\documentstyle[12pt] {article}
\def\zid{1\kern-0.36em\llap~1}

\newcommand{\beq}{\begin{equation}}
\newcommand{\ber}{\begin{eqnarray}}
\newcommand{\eeq}{\end{equation}}
\newcommand{\eer}{\end{eqnarray}}

\newcommand{\text}{}

\begin{document}

\begin{titlepage}
\rightline{SUNY BING 5/15/95}

\begin{center}
{\large \bf TESTS FOR TAU'S CHARGED-CURRENT
STRUCTURE}\\[2mm]
Charles A. Nelson\footnote{Electronic address: cnelson @
bingvmb.cc.binghamton.edu Contributed paper to LP95, Beijing.
\newline Revision of SUNY BING [10/1/94] which will appear in
Phys. Let. } \\
{\it Department of Physics\\
State University of New York at Binghamton\\
Binghamton, N.Y. 13902-6016}\\[5mm]
\end{center}

\vspace{2mm}
\begin{abstract}

The Lorentz structure of the tau lepton's charged-current can
almost
be completely
\newline determined  by use of stage-two spin-correlation
functions
for
the $\{\rho ^{-},\rho ^{+}\}$and $\{a_1^{-},a_1^{+}\}$ decay
modes.  It is
possible to test for a ``$(V-A)$ $ + $ something'' structure in the
${J^{Charged}}_{Lepton}$ current, so as to bound the scales
$\Lambda$ for ``new physics" such as arising from tau weak
magnetism,  weak
electricity, and/or second-class currents. In practice, only limited
information can
be obtained from the $\tau \rightarrow \pi \nu$ channels.
\end{abstract}

\end{titlepage}

\begin{center}
{\bf Text}
\end{center}

Based on the assumption of a mixture of V and A couplings in
the
$\tau$ charged-current, experiments at $e^- e^+$ colliders have
been
setting limits
on the presence
of $(V+A)$ couplings in $\tau ^{-}\rightarrow A^{-}\nu _\tau $
decay for
$A=a_1, \rho, \pi, (l \bar{\nu_l})$.  The mixture of V and A
couplings can be
characterized by the value of the ``chirality parameter'' $ \xi
_A\equiv
\frac{|g_L|^2-|g_R|^2}{|g_L|^2+|g_R|^2}=\frac{2Re\left(
v_Aa_A^{*}\right) }{%
|v_A|^2+|a_A|^2}$.   Note that $\xi _A=-\langle h_{\nu _\tau
}\rangle $, twice the
negative of the $\nu _\tau $ helicity, in the special case for a
spin-
one $A^{-
}$ particle of only $V$ and $A$
couplings and $m_\nu =0$. Using spin-correlations, the ARGUS
\cite{e1},
ALEPH \cite{e2}, and CLEO \cite{e3} collaborations have
measured $\xi _A$.
The current world average is $\xi
_A=1.002\pm 0.032$ \cite{e3}.  So the leading contribution
in the tau's charged-current is $(V-A)$ to better than the $5\%$
level.

Therefore, the focus of this paper is on tests for ``something'' in a
``$(V-A)$ $ + $
something'' structure in the tau's ${J^{Charged}}_{Lepton}$
current
\cite{e4} .
This extra contribution can
show up experimentally because of its interference with the $(V-
A)$
part which,
we assume, arises as predicted by the standard lepton model.
More
precisely, the
idea is to search for ``additional structure'' due to additional
Lorentz
couplings in
${J^{Charged}}_{Lepton}$ by generalizing the $\tau $ spin-
correlation function
$I(E_\rho ,E_{\bar B})$ by including the $\rho $ polarimetry
information
\cite{a2,c1} that is available from the $%
\rho^{ch}\rightarrow \pi^{ch}\pi ^o$ decay distribution
\cite{C94}.
The
symbol $B=\rho
,\pi ,l$ . Since this adds on spin-correlation information from the
next
stage of decays in the decay sequence, we call such an energy-
angular
distribution a stage-two spin-correlation (S2SC) function.
Similarly,
$a_1$
polarimetry information can be included from the  $\tau ^{-
}\rightarrow a_1^{-
}\nu \rightarrow \left( \pi ^{-}\pi
^{-}\pi ^{+}\right) \nu ,\left( \pi ^o\pi ^o\pi ^{-}\right) \nu $
decay
modes
\cite{C94a}.

The simplest useful S2SC is for the $CP$-symmetric decay
sequence
$Z^o$, or
$%
\gamma ^{*}\rightarrow \tau ^{-}\tau ^{+}\rightarrow (\rho ^{-
}\nu
_\tau
)(\rho ^{+}\bar \nu _\tau )$ followed by both $\rho ^{\mp
}\rightarrow \pi
^{\mp }\pi ^o$,
\ber
\text{I(E}_\rho \text{,E}_{\bar \rho }\text{,}\tilde \theta
_1\text{,}\tilde
\theta _{2\text{ }}\text{) = }|\text{T}\left( +-\right) |^2\rho
_{++}\bar
\rho _{--} +\text{ }|\text{T}\left( -+\right) |^2\rho _{--}\bar \rho
_{++} \nonumber  \\ +\text{ }|\text{T}\left( ++\right) |^2\rho
_{++}\bar \rho
_{++}
 +\text{ }|\text{T}\left( --\right) |^2\rho
_{--}\bar \rho _{--}
\eer
If we think in terms of probabilities, the quantum-mechanical
structure of this
expression is apparent,  since the $T(\lambda _{\tau
^{-}},\lambda _{\tau ^{+}})$ helicity amplitudes
describe the production of the $(\tau ^{-}\tau ^{+})$ pair via
$Z^o$,
or $%
\gamma ^{*}\rightarrow \tau ^{-}\tau ^{+}$. For instance, in the
1st
term,
the factor  $|T(+,-)|^2=$``Probability to produce a $\tau ^{-}$
with
$%
\lambda _{\tau ^{-}}=\frac 12$ and a $\tau ^{+}$ with $\lambda
_{\tau
^{+}}=-\frac 12$ '' is multiplied by the product of the decay
probablity, $\rho _{++}$, for the positive helicity $\tau ^{-
}\rightarrow
\rho ^{-}\nu \rightarrow \left( \pi ^{-}\pi ^o\right) \nu $ times the
decay
probablity, $\bar \rho _{--}$, for the negative helicity $\tau
^{+}\rightarrow \rho ^{+}\bar \nu \rightarrow \left( \pi ^{+}\pi
^o\right)
\bar \nu $ .

The kinematic variables in $I_4$ are the usual ``spherical" ones
which naturally
appear in
the helicity formalism in describing such a decay sequence.  The
1st
stage of the
decay sequence $\tau
^{-},  \tau ^{+}\rightarrow (\rho ^{-}\nu _\tau )(\rho ^{+}\bar \nu
_\tau )$ is
described by the 3 variables $\theta _1^\tau,  \theta _2^\tau ,  \cos
\phi  $ where
$\phi$ is the opening $\angle $ between the two decay planes.
These
are
equivalent
to the $Z^o$, or $\gamma ^{*}$
center-of-mass variables, $E_{\rho}, E_{\bar \rho }, \cos \psi $.
Here $\psi =$%
``opening $\angle $ between the $\rho ^{-}$ and $\rho
^{+}$momenta in the $%
Z/\gamma ^{*}$ cm''. When the Lorentz ``boost'' to one of the
$\rho
$ rest frames
is directly from the
$Z/\gamma
^{*}$ cm frame, the 2nd stage of the decay sequence is described
by
the
usual 2 spherical angles for the $\pi ^{ch}$ momentum
direction in that $\rho $ rest frame: $\tilde \theta _1, \tilde \phi
_1$
for $\rho _1^{-
}\rightarrow \pi _1^{-}\pi _1^o$, and $\tilde \theta _2, \tilde \phi
_2$
for $\rho
_2^{+}\rightarrow \pi _2^{+}\pi _2^o$. (See figures in
\cite{C94}.)

In (1), the composite decay density matrix elements are simply
the
decay
probability for a $\tau _1^{-}$ with
helicity $\frac h2$ to decay $\tau ^{-}\rightarrow \rho ^{-}\nu
\rightarrow
\left( \pi ^{-}\pi ^o\right) \nu $ since $
{d\text{N}}/{d\left( \cos \theta _1^\tau \right) d\left( \cos \tilde
\theta _1\right) }=\rho _{hh}\left( \theta _1^\tau ,\tilde \theta
_1\right)  $ and for the decay of the $\tau _2^{+}$ , $\bar \rho
_{hh}=\rho _{-h,-
h}\left( {\text{subscripts}} \quad  1
\rightarrow 2, a
\rightarrow
b \right)$.  For a $\tau _1^{-}$ with
helicity $\frac h2$ to decay $\tau ^{-}\rightarrow \rho ^{-}\nu _L
\rightarrow
\left( \pi ^{-}\pi ^o\right) \nu _L $
\ber
\rho _{hh}=
  \left( 1+h\cos \theta _1^\tau \right) \left[ \cos ^2\omega
_1\cos^2\tilde \theta
_1+\frac 12\sin ^2\omega _1\sin ^2\tilde \theta _1\right]
\nonumber
\\
+ \frac{r_a^2}2\left( 1-h\cos \theta _1^\tau \right)
\left[ \sin ^2\omega_1\cos ^2\tilde \theta _1   +\frac 12\left(
1+\cos
^2\omega
_1\right) \sin^2\tilde \theta _1 \right]   \nonumber \\
 +h\frac{r_a}{\sqrt{2}}\cos \beta _a\sin \theta
_1^\tau \sin 2\omega _1\left[ \cos ^2\tilde \theta _1-\frac 12\sin
^2\tilde
\theta _1\right]
\eer
with the Wigner rotation angle $\omega_1 =
\omega_1(E_{\rho})$,
\cite{C94}.
The dynamical parameters to be experimentally measured are the
polar parameters
$\beta _a=\phi
_{-1}^a-
\phi _0^a$,
$\beta
_b=\phi _1^b-\phi _0^b$, and $r_a={|A\left( -1,-\frac 12\right)
|}/{|A\left( 0,-\frac
12\right) |}$, $r_b={|B\left( 1,\frac 12\right) |}/{|B\left( 0,\frac
12\right) |}$.  In the
standard lepton model with a pure $(V-A)$ coupling, the
predicted
values are $\beta _{a,b}=0,r_{a,b}=\frac{\sqrt{2}m_\rho
}{E_\rho
+q_\rho }\simeq \sqrt{2}m_\rho /m_\tau \simeq 0.613.$  Note
that
the above $I_4$
spin-correlation function only depends on
4 of the above 7 kinematic variables.  Refs. \cite{C94,C94a} give
its
generalization,
$I_7$, which
also depends on
$\cos{\phi}, \tilde \phi _1,$ and $\tilde \phi _2 $.  We use $I_4$
in
this paper
because it is less complicated, has a useful sensitivity level, and
sometimes $I_7$
is not significantly better$^{\# 1}$.  \hskip 1em   \rm

For  the $\tau ^{-}\rightarrow
a_1^{-}\nu _L \rightarrow \left( \pi ^{-}\pi
^{-}\pi ^{+}\right) \nu _L ,\left( \pi ^o\pi ^o\pi ^{-}\right) \nu _L
$
modes,
\ber
\rho _{hh}=
 \left( 1+h\cos \theta _1^\tau \right) \left[ \sin ^2\omega _1\cos
^2\tilde \theta _1
+ ( 1- \frac 12\sin ^2\omega _1 ) \sin ^2\tilde \theta _1\right]
\nonumber \\
+ \frac{r_a^2}2\left( 1-h\cos \theta _1^\tau \right)   \left[ \left(
1+\cos
^2\omega
_1\right) \cos ^2\tilde \theta _1  +\left( 1+\frac 12\sin ^2\omega
_1\right) \sin
^2\tilde \theta _1 \right]  \nonumber  \\
-h\frac{r_a}{\sqrt{2}}\cos \beta _a\sin \theta
_1^\tau \sin 2\omega _1\left[ \cos ^2\tilde \theta _1-\frac 12\sin
^2\tilde
\theta _1\right]
\eer
Here $\tilde \theta _1$ specifies the normal to the $\left( \pi ^{-
}\pi
^{-}\pi ^{+}\right) $ decay triangle, instead of the $\pi ^{-}$
momentum
direction used for $\tau ^{-}\rightarrow \rho ^{-}\nu $.  The
Dalitz
plot for $\left( \pi ^{-}\pi ^{-}\pi ^{+}\right) $ has been
integrated
over
so that  \cite{bj} it is not necessary to separate the
form-factors for $a_1^{-} \rightarrow $ $\left( \pi ^{-}\pi ^{-}\pi
^{+}\right) $.

It is
straightforward to include $\nu _R$ and $\bar \nu _L$ couplings
in
$S2SC$
functions since
\ber
I\left( E_\rho ,E_{\bar \rho },\tilde \theta _1,\tilde \theta _2\right)
\mid
_{\nu _R,\bar \nu _L}=I_4 +\left( \lambda _R\right) ^2I_4\left(
\rho
\rightarrow
\rho ^R\right)
+\left( \bar \lambda _L\right) ^2I_4\left( \bar \rho \rightarrow \bar
\rho
^L\right) \nonumber  \\
+\left( \lambda _R\bar \lambda _L\right) ^2I_4\left( \rho
\rightarrow
\rho
^R,\bar \rho \rightarrow \bar \rho ^L\right)
\eer
where $\lambda _R\equiv {|A\left( 0,\frac 12\right) |}/{|A\left(
0,-\frac 12\right) |}$, $ \bar \lambda _L\equiv {|B\left( 0,-
\frac
12\right) |}/{|B\left( 0,\frac 12\right) |}$
give the moduli's of the $\nu _R$
and $\bar \nu _L$ amplitudes versus the standard amplitudes.
The
$\rho _{hh}$'s
for  $\tau \rightarrow \rho \nu $ with
$\nu _R$
and $\bar \nu _L$ final state particles are given by the
substitution
rules: $\rho _{hh}^R=\rho _{-h,-h}\left( r_a\rightarrow
r_a^R,\beta
_a\rightarrow
\beta _a^R\right)$,  $\bar \rho _{hh}^L=\bar \rho _{-h,-h}\left(
r_b\rightarrow
r_b^L,\beta
_b\rightarrow \beta _b^L\right)$.  The helicity amplitudes$^{\#2}$
\hskip 1em   \rm
for \hskip
1em  $\tau^{-
}\rightarrow \rho
^{-}\nu _{L,R}$ for both $(V\mp A)$ couplings and
$%
m_\nu $ arbitrary are given in \cite{C94}.

Historically in the study of the weak charged-current in muonic
and
in
hadronic processes, it has been important to determine the
``complete Lorentz structure'' directly from experiment. Here the
$I_4$ and $I_7$
functions can be
used to do this for the $ \tau$ charged-current since these
functions
depend directly on the 4 helicity amplitudes for $\tau
^{-}\rightarrow \rho ^{-}\nu $ and on the 4 amplitudes for the
$CP$-
conjugate
process. In this paper, for $I_4$ we report the associated ``ideal''
sensitivities. We first consider the ``traditional'' couplings for
$\tau
^{-
}\rightarrow \rho
^{-}\nu $ which characterize the most general Lorentz coupling $
\rho _\mu ^{*}\bar u_{\nu _\tau }\left( p\right) \Gamma ^\mu
u_\tau
\left(
k\right)  $, $k_\tau =q_\rho +p_\nu $. It is convenient to treat
separately the vector
and
axial vector matrix elements. We introduce a parameter
$%
\Lambda =$ ``the scale of New Physics''. In effective field theory
this
is the scale at which new particle thresholds are expected to
occur.  In old-fashioned renormalization theory it is the scale at
which the calculational methods and/or the principles of
``renormalization''
breakdown, see e.g. \cite{th}. While some terms of the
following types do
occur as higher-order perturbative-corrections in the standard
model,
such SM
contributions  are ``small'' versus the sensitivities of present tests
in
$\tau$ physics
in the analogous cases of the $\tau$'s neutral-current and
electromagnetic-current
couplings, c.f.
\cite{d0}.  For charged-current couplings, the situation should be
the
same.

In terms of the ``traditional'' tensorial and spin-zero couplings%
\ber
V_{\nu \tau }^\mu \equiv \langle \nu |v^\mu \left( 0\right) |\tau
\rangle
=\bar u_{\nu _\tau }\left( p\right) [g_V\gamma ^\mu
+\frac{f_M}{2\Lambda }\iota \sigma ^{\mu \nu }(k-p)_\nu
+\frac{g_{S^{-}}}{%
2\Lambda }(k-p)^\mu ]u_\tau \left( k\right) \\
A_{\nu \tau }^\mu \equiv \langle \nu |a^\mu \left( 0\right) |\tau
\rangle
=\bar u_{\nu _\tau }\left( p\right) [g_A\gamma ^\mu \gamma _5
+\frac{f_E}{2\Lambda }\iota \sigma ^{\mu \nu }(k-p)_\nu
\gamma
_5+\frac{%
g_{P^{-}}}{2\Lambda }(k-p)^\mu \gamma _5]u_\tau \left(
k\right)
\eer
\noindent
Notice that $\frac{f_M}{2\Lambda }=$ a ``tau weak magnetism''
type
coupling, and $\frac{f_E}{2\Lambda }=$ a ``tau weak
electricity''
type
coupling. Both the scalar $g_{S^{-}}$ and pseudo-scalar
$g_{P^{-
}}$%
couplings do not contribute for  $\tau ^{-}\rightarrow \rho ^{-
}\nu $
since $%
\rho _\mu ^{*}q^\mu =0$, nor for  $\tau ^{-}\rightarrow a_1^{-
}\nu
$.
By Lorentz invariance, there are the equivalence theorems for the
vector
current%
\ber
S\equiv V+f_M, & T^{+}\equiv -V+S^{-}
\eer
\noindent
and for the axial-vector current
\ber
P\equiv -A+f_E, & T_5^{+}\equiv A+P^{-}
\eer
where
\beq
\Gamma _V^\mu =g_V\gamma ^\mu +
\frac{f_M}{2\Lambda }\iota \sigma ^{\mu \nu }(k-p)_\nu   +
\frac{g_{S^{-}}}{2\Lambda }(k-p)^\mu +\frac{g_S}{2\Lambda
}(k+p)^\mu
+%
\frac{g_{T^{+}}}{2\Lambda }\iota \sigma ^{\mu \nu }(k+p)_\nu
\eeq
\beq
\Gamma _A^\mu =g_A\gamma ^\mu \gamma _5+
\frac{f_E}{2\Lambda }\iota \sigma ^{\mu \nu }(k-p)_\nu \gamma
_5
+
\frac{g_{P^{-}}}{2\Lambda }(k-p)^\mu \gamma
_5+\frac{g_P}{2\Lambda }%
(k+p)^\mu \gamma _5  +\frac{g_{T_5^{+}}}{2\Lambda }\iota
\sigma ^{\mu \nu
}(k+p)_\nu \gamma _5
\eeq
The matrix elements of the divergences of these charged-currents
are
\beq
(k-p)_\mu V^\mu =[g_V(m_\tau -m_\nu )
+
\frac{g_{S^{-}}}{2\Lambda }q^2+\frac{g_S}{2\Lambda
}(m_\tau
^2-m_\nu ^2)
 +%
\frac{g_{T^{+}}}{2\Lambda }(q^2-[m_\tau -m_\nu ]^2)]\bar
u_\nu
u_\tau
\eeq
\beq
(k-p)_\mu A^\mu =[- g_A(m_\nu +m_\tau )
+
\frac{g_{P^{-}}}{2\Lambda }q^2+\frac{g_P}{2\Lambda
}(m_\tau
^2-m_\nu ^2)
 +%
\frac{g_{T_5^{+}}}{2\Lambda }(q^2-[m_\tau+m_\nu ]^2)]\bar
u_\nu \gamma
_5u_\tau
\eeq
Both the weak magnetism  $\frac{f_M}{2\Lambda }$ and the
weak
electricty $%
\frac{f_E}{2\Lambda }$ terms are divergenceless. On the other
hand, since $%
q^2=m_\rho ^2$,  when $m_\nu =m_\tau $ there contributions
from
$S^{-
},T^{+},A,P^{-},T_5^{+}$.

Table 1 gives the limits on these additional couplings assuming a
``$(V-A)$ $+$%
something'' structure for the $\tau$ charged-current assuming real
coupling
constants.  At $M_Z$ the scale of $\Lambda \approx
$few $\
100GeV
$ can be probed; and at $10GeV$ or at $4GeV$ the scale of $1-
2TeV$ can be
probed.

The tables list only the ideal statistical errors \cite{c1},
and assume
respectively $10^7 Z^{o}$ events and $10^7$ $(\tau^{-} \tau^{+}
)$
pairs.
For the $\rho$ mode, we use B($\tau \rightarrow \rho \nu $) =
$24.6$\%. For the
$a_1$ mode we sum the charged plus neutral
pion $a_1$ final states so B($\tau \rightarrow {a_1}^{ch+neu}
\nu
$) = $18$\%,
and use $m_{a_1} = 1.275GeV$. The results in these tables
simply
follow by
using (7-8) and from the dependence
of the helicity amplitudes for $\tau ^{-}\rightarrow \rho ^{-}\nu $
on
the
presence of $(S\pm P)$ couplings with $m_\nu $ arbitrary:
\ber
A(0,-\frac 12) & =g_{S+P}(
\frac{m_\tau }{2\Lambda })\frac{2q_\rho }{m_\rho
}\sqrt{m_\tau
(E_\nu
+q_\rho )}  +g_{S-P}(\frac{m_\tau }{2\Lambda })\frac{2q_\rho
}{m_\rho }%
\sqrt{m_\tau (E_\nu -q_\rho )}, \quad
A(-1,-\frac 12) & =0
\eer
and
\ber
A(0,\frac 12) & =g_{S+P}(
\frac{m_\tau }{2\Lambda })\frac{2q_\rho }{m_\rho
}\sqrt{m_\tau
(E_\nu
-q_\rho )}  +g_{S-P}(\frac{m_\tau }{2\Lambda })\frac{2q_\rho
}{m_\rho }%
\sqrt{m_\tau (E_\nu +q_\rho )}, \quad
A(1,\frac 12) & =0
\eer

In compiling the entries in Table 1, we have adopted the idea of
1st
and 2nd
class currents \cite{sc1}. This is suggested by a 3rd-family
perspective of a possible ``$\tau \leftrightarrow \nu _\tau $
symmetry''
in the structure of the tau lepton currents. At the level of the
masses,
this truly is a badly broken symmetry$^{\# 2}$.

\noindent But heeding the precedent historical successes of the
SM
in regard to
current-versus-mass symmetry distinctions, we believe that this
symmetry might
nevertheless be relevant to 3rd-family currents. Therefore, we
assume that the
effective charged-current ${J_{Lepton}}^{Charged}$ is
Hermitian
and has such
an SU(2) symmetry, so that we can identify the $\nu _\tau $ and
the
$\tau ^{-}$
spinors. Thereby, we obtain for the ``traditional couplings'' and
real
form factors
that the ``Class I'' couplings are $V,A,f_M,P^{-}$, and that the
``Class
II'' couplings are  $f_E,S^{-}$ if we define  $J_{Lepton}^\mu =$
$J_I^\mu +$ $%
J_{II}^\mu $ where for $U=\exp (\iota \pi I_2)$%
$$
\begin{array}{cc}
(J_I^\mu )^{\dagger }=-UJ_I^\mu U^{-1} & First \\
(J_{II}^\mu )^{\dagger }=UJ_{II}^\mu U^{-1} & Second
\end{array}
Class
$$

This classification is particularly useful in considering the reality
structure of the charged-current \cite{sc2}. As show in Table 2
there
is a ``clash''
between the ``Class I and Class II'' structures and
the consequences of
time-reversal invariance. In particular, there are the useful
theorems
that (a) ($\tau \leftrightarrow \nu _\tau $ symmetry) + ($T$
invariance) $%
\Longrightarrow $ Class II currents are absent, (b) ($\tau
\leftrightarrow
\nu _\tau $ symmetry) + (existence of $J_I^\mu $ and
$J_{II}^\mu
$) $%
\Longrightarrow $ violation of $T$ invariance, and (c) (existence
of
$%
J_{II}^\mu $) +  ($T$ invariance) $\Longrightarrow $($\tau
\leftrightarrow \nu _\tau $ symmetry) in $J_{Lepton}^\mu $ is
broken.

Table 3 shows the limits on such couplings assuming a pure-
imaginary
coupling constant. In the case of $(V-A)$ the limits on the $ \beta
$'s
in Refs.
\cite{C94,C94a,e4} cover this situation. The limits here are
in $ ( \Lambda
)^2$ with $\Lambda \sim$ few $10GeV$'s because this is
not a S2SC interference effect.

Besides the 3rd-family perspective of a possible $\tau
\leftrightarrow
\nu _\tau $ symmetry, it is also instructive to consider
``additional
structure'' in the
$\tau$ charged-current from the
viewpoint of ``Chiral Combinations'' of the various
Lorentz couplings.  This is especially interesting because the
$S\pm
P$ couplings
do not contribute to the transverse $\rho $ or $a_{1}$ transitions.
Tables 4 and 5
give the
limits on $\Lambda $ in the case of purely real and
imaginary coupling constants for the ``Chiral Couplings''.

Finally, Table 6,   the helicity amplitudes themselves
provide a simple framework for
characterizing a ``complete measurement'' of $\tau ^{-
}\rightarrow
\rho
^{-}\nu $ and of $\tau ^{-}\rightarrow a_1^{-}\nu $: For either,
when only $\nu _L$ coupling's exist, there are only 2
amplitudes, so 3 measurements,  of $r_a,\beta _a,$and
$%
|A(0,-\frac 12)|$ via $\{\rho ^{-},B^{+}\}\mid _{B\neq \rho }$,
will
provide
a ``complete measurement''. When $\nu _R$ coupling's also exist,
then
there are 2 more amplitudes, $A(0,\frac 12)$ and $A(1,\frac 12)$.
Then to
achieve
an ``almost'' complete measurement, 3 additional quantities must
be
determined, e.g. by the $I_4$:  $r_a^R,\beta _a^R$
and $%
\lambda _R\equiv \frac{|A(0,\frac 12)|}{|A(0,-\frac 12)|}$. But to
also
measure
the relative phase of the $\nu _L$ and $\nu _R$ amplitudes,
$\beta
_a^o\equiv
\phi _o^{aR}-\phi _o^a$ or $\beta _a^1\equiv \phi _1^a-\phi _{-
1}^a$, requires,
e.g., the occurrence of a common final state which arises from
both
$\nu _L$ and
$\nu _R$.

For comparison, Table 7 shows what can be learned from the
$\tau^{\pm}
\rightarrow \pi^{\pm} \nu$ decay {mode$^{\# 3}$}. \hskip  1em
\rm
The  \newline
\hskip0pt $\xi_\pi$ parameter and the $\Gamma (\tau \rightarrow
\pi \nu )$ partial width are the only observables for these modes.
While the
$f_M(q^2)$  and  $f_E(q^2)$ couplings do not contribute
to this decay mode, useful bounds can be obtained for the
$V+A$,
$S+P$, and
$T^{+} + {T_5}^{+}$ chiral couplings.  In principle this channel
is
particularly
important for the $S^{-} \pm P^{-}$ couplings contribute here
whereas they do
not for the $\rho$ and $a_{1}$ modes.  However, in the
$\tau^{\pm} \rightarrow \pi^{\pm} \nu$ decay amplitudes, each
such
coupling
appears
multiplied by a suppression factor of ${m_{\pi}}^2 /
({m_{\tau}}^2
-
{m_{\nu}}^2)$.  Hence, one conclusion of this paper is that both
the
present and
potential experimental bounds on $S^{-} \pm P^{-}$ couplings
are
exceptionally
poor or non-existent from measurements of the $\pi$, $\rho$ and
$a_{1}$ modes
in tau lepton decays!

In conclusion,  $(\tau ^{-}\tau ^{+})$ spin correlations$^{\# 4}$
\hskip
1em  \rm
with  \hskip 1em  $\rho$ and $a_1$ polarimetry observables can
be
used to probe
for
``additional structure'' in the tau's charged-current. For example,
tau weak magnetism, $f_M(q^2)$,  and  tau weak
electricity, $f_E(q^2)$, can be probed to new physics scales of
\quad$\Lambda _{RealCoupling} \sim  1.2-1.5TeV$ \rm at $10$,
or
$4GeV$ and
\quad$\Lambda _{Imag.Coupling} \sim 28-34GeV$ \rm at $10,$
or
$4GeV$. \rm
By spin-correlation techniques the Lorentz stucture of the $\tau$
charge-current
can almost be completely determined from the $\{\rho ^{-},\rho
^{+}\}$and $%
\{a_1^{-},a_1^{+}\}$ modes.

\section*{Acknowledgments}
For helpful discussions, we thank physicists at
Cornell,
DESY, Valencia, and at Montreux. This work was
partially
supported by U.S. Dept. of Energy Contract No. DE-FG 02-
96ER40291.

\section*{Footnotes}
\begin{enumerate}
\item  For testing for $(V+A)$ versus  $(V-A)$, the use of  $I_7$
for $\{\rho
^{-},\rho ^{+}\}$ gives less than a $1\%$ improvement over
$I_4$
at $M_Z$, $10
GeV$, or $4 GeV$.  If in addition the $\tau ^{-}$momentum
direction is
known via a SVX detector, there is only an $\sim 11\%$
improvement.  The same
numbers occur for $\{ {a_1}^{-
},{a_1}^{+} \}$.  In contrast, by using $I_4$, instead of the
simpler 2 variable $I\left( E_\rho ,E_{\bar \rho }\right) $ spin-
correlation
function, there is about a factor of 8 improvement at $M_Z$.
\item  Note $\frac{m_b}{m_t}\sim \frac 5{174}\sim 3\%$, and
$%
\frac{m_\nu }{m_\tau }<\frac{23.8}{1777}\sim 1.4\%$ so this
symmetry is badly
broken in the masses for the 3rd family. However, for the other
leptons this
symmetry may be more strongly broken since
$\frac{m_{\nu
_e}}{m_e}<10^{-5}$, and $\frac{m _{\nu _\mu }}{m_\mu
}<0.15\%$ from the
current empirical bounds.  From phenomenological mass
formulas,
e.g. \cite{har}, such as the GUT mass
formula, $\nu_{\tau}$:$\nu_{\mu}$:$\nu_{e}$ $\sim$
${m_t}^2$:${m_c}^2$:${m_u}^2$, the tau leptons are also the
least
asymmetric
since then $\frac{m_{\nu
_\tau}}{m_\tau} \approx 10^{-8}$, $\frac{m_{\nu
_\mu}}{m_\mu} \approx 10^{-11}$, and $\frac{m_{\nu
_e}}{m_e} \approx 3\cdot 10^{-14}$ for the normalization
$m_{\nu_{\tau}}=20eV$.
\item  Details on the analysis of the $\tau \rightarrow \pi \nu$
modes
will be
reported elsewhere \cite{C94a}.
\item  The tests in this paper use ($\tau^{-} \tau^{+}$) spin-
correlations as it is assumed that the $e^{-}$ and $e^{+}$
colliding
beams are
not
longitudinally-polarized.  Recently, Y.-S. Tsai
\cite{ch1,ch2} has shown that in tau decays the sensitivities of
tests
for $CP$
violation, and
for other types of ``new physics'',  are substantially improved in
regard to both
systematic and statistical errors by the use of longitudinally-
polarized beams at the
($\tau^{-} \tau^{+}$) threshold.
\end{enumerate}

\section*{Table Captions}
\quad Table 1: Limits on $\Lambda$ in $GeV$ for Real $g_i$'s.
For
$V+A$ only, the entry is for $\xi_A$.

Table 2: ``Reality structure'' of $J^{\mu}_{Lepton}$ current's
form
factors.

Table 3: Limits on $\Lambda$ in $GeV$ for Pure Imaginary
$g_i$'s.
For $V+A$
only, the entry is for $\xi_A$.

Table 4: ``Chiral Couplings'':  Limits on $\Lambda$ in $GeV$
for
Real $g_i$'s. For the $\rho$ and $a_1$ modes, equivalent
couplings
are
$T^{+}+T_5^{+} \equiv V-A$; $T^{+}- T_5^{+}\equiv V+A$.

Table 5: ``Chiral Couplings'':  Limits on $\Lambda$ in $GeV$
for
Pure Imaginary $g_i$'s. For the $\rho$ and $a_1$
modes, equivalent couplings are
$T^{+}+T_5^{+}\equiv V-A$; $T^{+}-T_5^{+}\equiv V+A$.

Table 6: Elements of error matrix for limits on $\nu_R$ and
$\overline{\nu}_L$
couplings in terms of  the
helicity amplitudes for respectively $\tau \rightarrow \rho \nu$,
and
$\tau
\rightarrow a_1 \nu$.

Table 7: ``Chiral Couplings'': Limits on $\Lambda$ in $GeV$
from
$\tau
\rightarrow \pi \nu$.  $\xi_\pi$ entries are  for $\lbrace \pi^{-},
\pi^{+} \rbrace$
spin correlations at $M_Z$, $10GeV$, $4GeV$.  $\Gamma (\tau
\rightarrow \pi \nu )$ entries follow from current data \cite{mt}.

\newpage

\begin{table*}[h]
\setlength{\tabcolsep}{1.5pc}
\newlength{\digitwidth} \settowidth{\digitwidth}{\rm 0}
\catcode`?=\active \def?{\kern\digitwidth}
\caption{}
\label{tab1}
\begin{tabular*}{\textwidth}{@{}l@{\extracolsep{\fill}}rrrr}
\hline
                 & \multicolumn{2}{l}{$\lbrace \rho^{-}, \rho^{+}
\rbrace$ mode}
                 & \multicolumn{2}{l}{$\lbrace a_{1}^{-},
a_{1}^{+}
\rbrace$ mode}
\\
\cline{2-3} \cline{4-5}
                 & \multicolumn{1}{r}{At $M_Z$}
                 & \multicolumn{1}{r}{10, or 4 GeV}
                 & \multicolumn{1}{r}{At $M_Z$}
                 & \multicolumn{1}{r}{10, or 4 GeV}         \\
\hline
{\bf 1st Class Currents}  &                      &                   &
&               \\
$V+A$, for $\xi_A$        & $0.006$      & $0.0012$ & $0.010$
&
$0.0018$ \\
$f_M$, for $\Lambda$   & $214 GeV$ & $1,200$   &  $282$   &
$1,500$ \\
$S$                                   & $306 GeV$ &  $1,700$   &  $64$
&
$345$ \\
$T_5^{+}$                      & $506 GeV$ &  $2,800$   &  $371$
&
$2,000$ \\
{\bf 2nd Class Currents}&                      &                   &
&
\\
$f_E$, for $\Lambda$    & $214 GeV$ & $1,200$    &  $282$
&
$1,500$ \\
$P$                                   & $306 GeV$ &  $1,700$   &  $64$
&
$345$ \\
$T^{+}$                          & $506 GeV$ &  $2,800$   &  $371$
&
$2,000$ \\
\hline
\multicolumn{5}{@{}p{120mm}}{}
\end{tabular*}
\end{table*}

\begin{table*}[h]
\setlength{\tabcolsep}{1.5pc}
\caption{}
\label{tab2}
\begin{tabular*}{\textwidth}{@{}l@{\extracolsep{\fill}}|rr|r}
\hline
 {\bf Form Factor:}             & Class I Current & Class II
Current&
{\bf T
invariance} \\
\hline
$V,A,f_{M},P^{-}$           & Real parts          & Imaginary parts
&
$Re\neq 0,
Im=0$  \\
$f_{E},S^{-}$                    & Imaginary parts & Real parts
&
$Re\neq 0,
Im=0$  \\
\hline
\end{tabular*}
\end{table*}

\begin{table*}[hbt]
\setlength{\tabcolsep}{1.5pc}
\caption{}
\label{tab3}
\begin{tabular*}{\textwidth}{@{}l@{\extracolsep{\fill}}rrrr}
\hline
                 & \multicolumn{2}{l}{$\lbrace \rho^{-}, \rho^{+}
\rbrace$ mode}
                 & \multicolumn{2}{l}{$\lbrace a_{1}^{-},
a_{1}^{+}
\rbrace$ mode}
\\
\cline{2-3} \cline{4-5}
                 & \multicolumn{1}{r}{At $M_Z$}
                 & \multicolumn{1}{r}{10, or 4 GeV}
                 & \multicolumn{1}{r}{At $M_Z$}
                 & \multicolumn{1}{r}{10, or 4 GeV}         \\
\hline
{\bf 1st Class Currents:}     &                          &                  &
&
\\
$V+A$, for $\xi_A$             & $0.006$          & $0.0012$ &
$0.010$
& $0.0018$
\\
$f_M$, for $(\Lambda)^2$ & $(12GeV)^2$ & $(28)^2$ &
$(15)^2$& $(34)^2$ \\
$S$                                         & $(14GeV)^2$ & $(33)^2$ & $(
6)^2$ & $(13)^2$
\\
$T_5^{+}$                            & $(22GeV)^2$ & $(50)^2$ &
$(18)^2$& $(42)^2$
\\
{\bf 2nd Class Currents:}     &                          &                  &
&
\\
$f_E$, for $(\Lambda)^2$   & $(12GeV)^2$ & $(28)^2$ &
$(15)^2$& $(34)^2$ \\
$P$                                         & $(14GeV)^2$ & $(33)^2$ & $(
6)^2$ & $(13)^2$
\\
$T^{+}$                                 & $(22GeV)^2$ & $(50)^2$ &
$(18)^2$& $(42)^2$
\\
\hline
\multicolumn{5}{@{}p{120mm}}{}
\end{tabular*}
\end{table*}

\begin{table*}[hbt]
\setlength{\tabcolsep}{1.5pc}
\caption{}
\label{tab4}
\begin{tabular*}{\textwidth}{@{}l@{\extracolsep{\fill}}rrrr}
\hline
                 & \multicolumn{2}{l}{$\lbrace \rho^{-}, \rho^{+}
\rbrace$ mode}
                 & \multicolumn{2}{l}{$\lbrace a_{1}^{-},
a_{1}^{+}
\rbrace$ mode}
\\
\cline{2-3} \cline{4-5}
                 & \multicolumn{1}{r}{At $M_Z$}
                 & \multicolumn{1}{r}{10, or 4 GeV}
                 & \multicolumn{1}{r}{At $M_Z$}
                 & \multicolumn{1}{r}{10, or 4 GeV}         \\
\hline
$V+A$, for $\xi_A$                    & $0.006$         & $0.0012$ &
$0.010$ &
$0.0018$ \\
$S+P$, for $\Lambda$                & $310 GeV$   & $1,700$   &
$
64$     & $
350$ \\
$S-P$, for $(\Lambda)^2$          &$(11GeV)^2$& $(25)^2$ &
$(4)^2$  &
$(7)^2,(10)^2$ \\
$f_M+f_E$, for $\Lambda$      & $210 GeV$   & $1,200$   &
$280$   & $1,500$
\\
$f_M-f_E$, for $(\Lambda)^2$&$(9GeV)^2$  & $(20)^2$ &
$(10)^2$& $(24)^2$
\\
\hline
\multicolumn{5}{@{}p{120mm}}{}
\end{tabular*}
\end{table*}

\begin{table*}[hbt]
\setlength{\tabcolsep}{1.5pc}
\caption{}
\label{tab5}
\begin{tabular*}{\textwidth}{@{}l@{\extracolsep{\fill}}rrrr}
\hline
                 & \multicolumn{2}{l}{$\lbrace \rho^{-}, \rho^{+}
\rbrace$ mode}
                 & \multicolumn{2}{l}{$\lbrace a_{1}^{-},
a_{1}^{+}
\rbrace$ mode}
\\
\cline{2-3} \cline{4-5}
                 & \multicolumn{1}{r}{At $M_Z$}
                 & \multicolumn{1}{r}{10, or 4 GeV}
                 & \multicolumn{1}{r}{At $M_Z$}
                 & \multicolumn{1}{r}{10, or 4 GeV}         \\
\hline
$V+A$, for $\xi_A$                    & $0.006$         & $0.0012$ &
$0.010$ &
$0.0018$ \\
$S+P$, for$(\Lambda)^2$          &$(11GeV)^2$& $(25)^2$ &
$(4)^2$  &
$(10)^2$ \\
$S-P$, for $(\Lambda)^2$          &$(11GeV)^2$& $(25)^2$ &
$(4)^2$  &
$(7)^2,(10)^2$ \\
$f_M+f_E$, for$(\Lambda)^2$&$(9GeV)^2$  & $(20)^2$ &
$(10)^2$& $(24)^2$
\\
$f_M-f_E$, for $(\Lambda)^2$&$(9GeV)^2$  & $(20)^2$ &
$(10)^2$& $(24)^2$
\\
\hline
\multicolumn{5}{@{}p{120mm}}{}
\end{tabular*}
\end{table*}

\begin{table*}[hbt]
\setlength{\tabcolsep}{1.5pc}
\caption{}
\label{tab7}
\begin{tabular*}{\textwidth}{@{}l@{\extracolsep{\fill}}rrrr}
\hline
                 & \multicolumn{2}{l}{$\lbrace \rho^{-}, \rho^{+}
\rbrace$ mode}
                 & \multicolumn{2}{l}{$\lbrace a_{1}^{-},
a_{1}^{+}
\rbrace$ mode}
\\
\cline{2-3} \cline{4-5}
                 & \multicolumn{1}{r}{At $M_Z$}
                 & \multicolumn{1}{r}{10, or 4 GeV}
                 & \multicolumn{1}{r}{At $M_Z$}
                 & \multicolumn{1}{r}{10, or 4 GeV}         \\
\hline
{\bf Diagonal elements:}             &                     &
&
                       &    \\
$ a=\lambda_R$                           &(8\%)$^2$    &$(4\%)^2$
&
$(18\%)^2$ &$(8\%)^2,(9\%)^2$\\
$ b=\lambda_Rr_a^R$                &$(8\%)^2$   &$(4\%)^2$
&
$(18\%)^2$ &$(8\%)^2,(9\%)^2$\\
$ c = $                                          &                     &
&
                      &  \\
$(\lambda_R)^2r_a^R
\cos{\beta_a^R}$
&$(13\%)^2$&$(6\%)^2,(10\%)^2$&$(41\%)^2$
                      &$(20\%)^2,(24\%)^2$ \\
{\bf Correlations:}                     &                     &
&
                      &    \\
$ \rho_{ab}$                               &$-0.75$       &$-0.77$
&$-0.95$
                      &$-0.96, -0.97$ \\
$ \rho_{ac}$                               &$-0.27$       &$-0.17,0.06$
&$-0.56$
                      &$0.029,0.019$ \\
$ \rho_{bc}$                               &$0.085$      &$.017,0.003$
& $0.04$
                      &$-0.41,-0.026$ \\
\hline
\multicolumn{5}{@{}p{120mm}}{}
\end{tabular*}
\end{table*}

\begin{table*}[hbt]
\setlength{\tabcolsep}{1.5pc}
\caption{}
\label{tab7}
\begin{tabular*}{\textwidth}{@{}l@{\extracolsep{\fill}}rrr}
\hline
                 & \multicolumn{1}{c}{From $\xi_\pi$:}
                 & \multicolumn{2}{l}{From $\Gamma (\tau
\rightarrow
\pi \nu )$ }
 \\
\cline{2-4}
                 & \multicolumn{1}{c}{ $\vert g_i /  g_L \vert^2$ }
                 & \multicolumn{1}{r}{$\vert g_i /  g_L \vert^2 $}
                 & \multicolumn{1}{r}{$2 Re( {g_L}^* g_i )$}         \\
\hline
$V+A$, for $\xi_\pi$                                   &
$0.015,0.004,0.009$
&
$0.014$             & $            $ \\
$S+P, T^+ + {T_5}^+$, for$\Lambda$     &$           $
&
$     $                  & $127GeV$ \\
$S-P, T^+ -
{T_5}^+$,for$(\Lambda)^2$&$(10GeV)^2,(21GeV)^2,(13GeV)
^2$
&
$(<1GeV)^2$   & $   $ \\
$S^{-} +P^{-}$, for$\Lambda$                 &$           $
&
$     $                  & $<1GeV$ \\
$S^{-} -P^{-}$,for$(\Lambda)^2$           &$(<
1GeV)^2,(1.6GeV)^2,(1GeV)^2$&
$(<1GeV)^2$   & $   $ \\

\\
\hline
\multicolumn{4}{@{}p{100mm}}{}
\end{tabular*}
\end{table*}

\end{document}